\documentstyle[prl,tighten,aps,epsfig]{revtex}

\begin{document}

\title{Strong field limit of black hole gravitational lensing}

\author{V. Bozza\thanks{%
E-mail valboz@sa.infn.it}, S. Capozziello\thanks{E-mail
capozziello@sa.infn.it}, G. Iovane\thanks{E-mail
geriov@sa.infn.it}, G. Scarpetta\thanks{E-mail
scarpetta@sa.infn.it} \\ {\em Dipartimento di Fisica ``E.R.
Caianiello'', Universit\`a di Salerno, Italy.}\\
 {\em  Istituto Nazionale di Fisica Nucleare, Sezione di
 Napoli.}}

 \date{today}
\maketitle

\begin{abstract}
We give the formulation of the gravitational lensing theory in the
strong field limit for a Schwarzschild black hole as a counterpart
to the weak field approach. It is possible to expand the full
black hole lens equation to work a simple analytical theory that
describes at a high accuracy degree the physics in the strong
field limit. In this way, we derive compact and reliable
mathematical formulae for the position of additional critical
curves, relativistic images and their magnification, arising in
this limit.
\end{abstract}

\vspace{20. mm} PACS numbers: 95.30.Sf, 04.70.-s, 98.62.Sb

\section{Introduction}

The simplicity of the theory of gravitational lensing derives from
some basic assumptions which are satisfied by most physical
situations \cite{SEF}. Through the weak field and the thin lens
approximations, the whole theory was built and succeeded in
explaining the phenomenology risen up to now.

Yet, gravitational lensing must not be conceived as a weak field
phenomenon, since high bending and looping of light rays in strong
fields is one of the most well-known and amazing predictions of
general relativity. The importance of gravitational lensing in
strong fields is highlighted by the possibility of testing the
full general relativity in a regime where the differences with
non-standard theories would be manifest, helping the
discrimination among the various theories of gravitation
\cite{will}. However, the complexity of the full mathematical
treatment and the difficulties of experimental or observational
evidences are obstacles for these studies.

Without the linear approximations, we have trascendent equations
which are hard to be handled even numerically.

Several studies about light rays close to the Schwarzschild
horizon has been lead: for example, Viergutz \cite{Vie} made a
semi--analytical investigation about geodesics in Kerr geometry;
in refs. \cite{Bar,FMA} the appearance of a black hole in front of
a uniform background was studied. Recently, Virbhadra \& Ellis
\cite{VirEll} faced the simplest strong field problem, represented
by deflection in Schwarzschild space--time, by numerical
techniques. The existence of an infinite set of relativistic
images has been enlightened and the results have been applied to
the black hole at the centre of the Galaxy. Later on, by an
alternative formulation of the problem, Frittelli, Kling \& Newman
\cite{FKN} attained an exact lens equation, giving integral
expressions for its solutions, and compared their results to those
by Virbhadra \& Ellis.

All these studies are affected by the high complexity in the
investigation of the geodesics in the neighbourhood of the horizon
and must resort to numerical methods to return valuable results.
This issue prevents any kind of general and systematic
investigation of light bending in this region, since fair
analytical formulae for the interesting quantities are absolutely
missing.

Starting from the black hole lens equation of ref. \cite{VirEll},
we perform a set of expansions exploiting the
source--lens--observer geometry and the properties of highly
deflected light rays. In this way, we manage to solve the lens
equation and find analytical expressions for the infinite set of
images formed by the black hole. This approach leads to extremely
simple formulae which allow an immediate comprehension of the
problem and a straightforward application to the physically
interesting situations. This strong field approach can be surely
considered as the direct counterpart of the weak field limit for
its striking simplicity and reliability. On the other hand, this
approach allows to investigate where  the strong field limit
effects become relevant.

This paper is structured as follows: In Sect. 2, the black hole
lens equation is presented and the first basic approximations are
explained. Sect. 3 contains the calculation of the deflection
angle. In Sect. 4 the deflection angle is then plugged in the
black hole lens equation to derive the relativistic images, their
amplification and the critical curves. Sect. 5 contains a
discussion of these results. Conclusions are drawn in Sect. 6.

\section{The black hole lens equation}

\begin{figure*}
\resizebox{9cm}{!}{\includegraphics{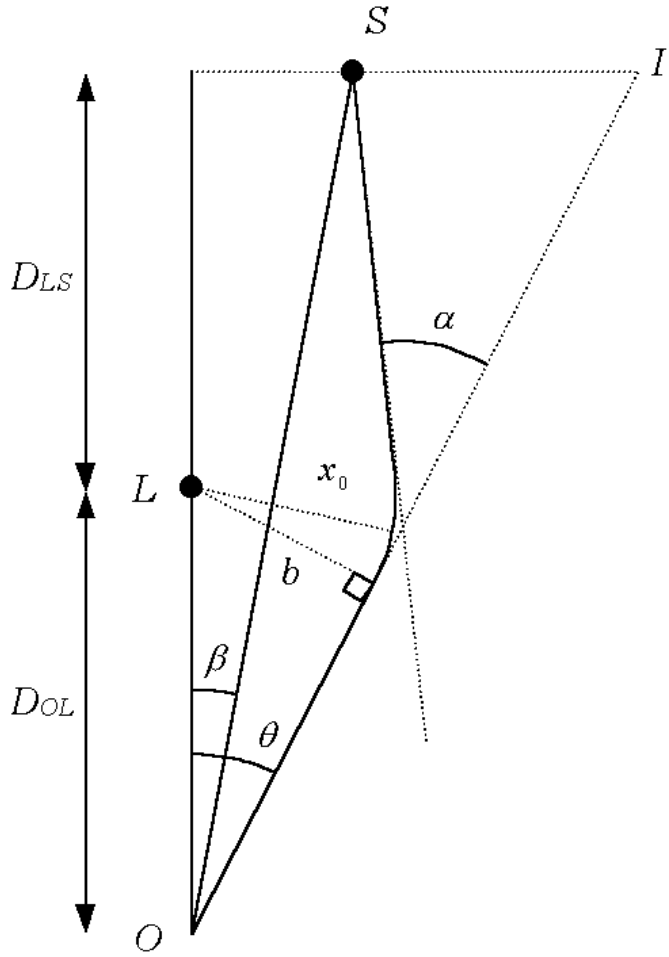}}
 \caption{Geometrical configuration of gravitational lensing. %
 Light rays emitted by the source $S$ are deflected by the lens $L$ %
 and reach the observer $O$ with an angle $\theta$, instead of $\beta$. %
 The total deflection angle is $\alpha$. $x_0$ is the closest approach distance %
 and $b$ is the impact parameter. $D_{OL}$ is the distance between the lens %
 and the observer. $D_{LS}$ is the distance between the lens %
 and the projection of the source on the optical axis $OL$. $D_{OS}=D_{OL}+D_{LS}$.}
 \label{F1}
\end{figure*}

The geometrical configuration of gravitational lensing is shown in
Fig. \ref{F1}. The light emitted by the source $S$ is deviated by
the black hole $L$ and reaches the observer $O$. Here, as a black
hole, we mean any compact object having a radius comparable to its
Schwarzschild radius, so that even very compact object which has
not undergone a full gravitational collapse would work in the same
way; $\beta$ is the angular position of the source with respect to
the optical axis $OL$ and $\theta$ is the angular position of the
image seen by the observer. It is important to stress that the
closest approach distance $x_0$ does not coincide with the impact
parameter $b$, unless in the limit of vanishing deflection angle
$\alpha$.

By inspection of Fig. \ref{F1}, it is possible to write a relation
among the source position, the image position and the deflection
angle $\alpha$.
\begin{equation}
\tan\beta=\tan\theta-\frac{D_{LS}}{D_{OS}}\left[ \tan\theta +\tan
\left(\alpha-\theta \right) \right] .%
\label{Full lens equation}
\end{equation}
This is what is called the full lens equation. Given a source
position $\beta$, the values of $\theta$, solving this equation,
give the position of the images observed by $O$.

In the weak field limit, several standard approximations are
performed. The tangents are expanded to the first order in the
angles since they are, at most, of the order of arcsec. The weak
field assumption reduces the deflection angle to ${\displaystyle
\frac{4GM}{c^2 x_0}}$. Then the lens equation can be solved
exactly and two images are found: one on the same side of the
source and one on the opposite. Their separations from the
optical axis are of the order of the Einstein angle
\begin{equation}
\theta_{E}=\sqrt{\frac{4GM}{c^2} \frac{D_{LS}}{D_{OS}D_{OL}}}.
\end{equation}
It is easy to see that, in most relevant cases, these images are
formed by light rays passing very far from the event horizon,
justifying the weak field approximation \cite{SEF}.

Now we turn to the study of gravitational lensing in strong field.
From now on, all lengths will be expressed in units of the
Schwarzschild radius ${\displaystyle \frac{2GM}{c^2}}$. The
deflection angle $\alpha$ contains the physical information about
the deflector and must be calculated through the integration of
the geodesic of the light ray \cite{Wei}. Its integral expression,
as a function of $x_0$, is
\begin{equation}
\alpha \left( x_0 \right) = \int\limits_{x_0}^\infty \frac{2}{x
\sqrt{\left( \frac{x}{x_0} \right)^2 \left( 1-\frac{1}{x_0}
\right) -\left( 1-\frac{1}{x} \right)}}{d}x-\pi .%
\label{Deflection angle}
\end{equation}
The next section is completely devoted to the achievement of a
manageable expression for this quantity as a function of $\theta$.

When the light ray trajectory gets closer to the event horizon,
the deflection increases. At some impact parameter, $\alpha$
becomes higher than $2 \pi$, resulting in a complete loop of the
light ray around the black hole. Decreasing further the impact
parameter, the light ray winds several times before emerging.
Finally, for $ b=(3 \sqrt{3})/2$, corresponding to $ x_0=3/2$, the
deflection angle diverges and the light ray is captured by the
black hole. For each loop we add to the light ray geodesic, there
is one particular value of the impact parameter such that the
observer is reached by the light coming from the source. So there
will be an infinite sequence of images on each side of the lens.

We shall put our attention on situations where the source is
almost perfectly aligned with the lens. In fact, this is the case
where the relativistic images are most prominent. In this case, we
are allowed to expand $\tan \beta$ and $\tan \theta$ to the first
order. Some more words are needed for the term $\tan \left(
\alpha- \theta \right)$. Even if $\theta$ is small, $\alpha$ is
not small in the situations of our interest. However, if a ray of
light emitted by the source $S$ is going to reach the observer
after turning around the black hole, $\alpha$ must be very close
to a multiple of $2 \pi$. Writing $\alpha=2n \pi+\Delta \alpha_n$,
with $n$ integer, we can perform the expansion $\tan \left(
\alpha- \theta \right) \sim \Delta \alpha_n -\theta$. The lens
equation becomes
\begin{equation}
\beta=\theta-\frac{D_{LS}}{D_{OS}} \Delta \alpha_n .%
\label{Small angles lens equation}
\end{equation}

As defined by (\ref{Deflection angle}), $\alpha$ is a positive
real number, corresponding to a clockwise winding in Fig.
\ref{F1}. We have inserted it in (\ref{Small angles lens
equation}) without giving a sign. Taking a positive $\beta$, this
equation  describes only images on the same side of the source
($\theta>0$). To obtain the images on the opposite side, we can
solve the same equation with the source placed in $-\beta$. Taking
the opposite of these solutions, we obtain the full set of the
secondary images.

\section{The deflection angle}

The deflection angle can be evaluated exactly, but its expression
does not allow the resolution of (\ref{Small angles lens
equation}). Anyway, it is possible to make some simplifying but
very general approximations which reduce the deflection angle to
an expression easier to handle. By this strategy,  we shall
calculate this fundamental quantity. All approximations are
essentially based on the proximity of the closest approach
distance $x_0$ to its minimum value which is $3/2$.

The integral in Eq.(\ref{Deflection angle}) gives the following
result
\begin{equation}
\alpha=-\pi-4 F \left( \phi_0,\lambda \right)G\left(x_0\right)
\end{equation}
where
\begin{equation}
G\left(x_0\right)=\sqrt{\frac{x_0 \left(
-3+3x_0-\sqrt{-3+2x_0+x_0^2}\right)}{ \left(3-2x_0 \right) \left(
1-x_0+\sqrt{-3+2x_0+x_0^2}\right)}}
\end{equation}
and
\begin{equation}
F \left( \phi_0,\lambda
\right)=\int\limits_0^{\phi_0}\left(1-\lambda \sin^2 \phi
\right)^{-1/2} {d} \phi%
 \label{Elliptic function}
\end{equation}
is an elliptic integral of first kind. The parameters $\phi_0$ and
$\lambda$ themselves are functions of $x_0$:
\begin{equation}
\phi_0=\arcsin \sqrt{\frac{-3+x_0-\sqrt{-3+2x_0+x_0^2}}
{2\left(-3+2x_0 \right)}} ,
\end{equation}
and
\begin{equation}
\lambda=\frac{3-x_0-\sqrt{-3+2x_0+x_0^2}}
{3-x_0+\sqrt{-3+2x_0+x_0^2}} .
\end{equation}

\begin{figure}
\resizebox{\hsize}{!}{\includegraphics{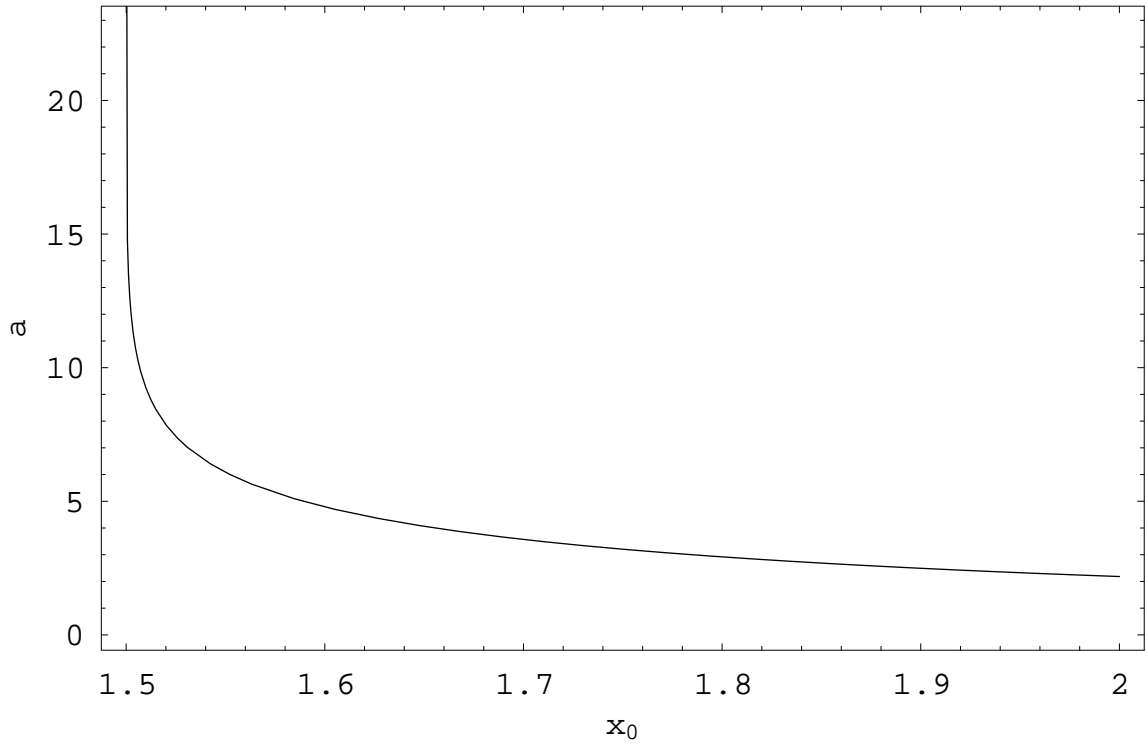}}
 \caption{Deflection angle as a function of the closest %
 approach distance.}
 \label{F2}
\end{figure}

Fig. \ref{F2} shows a plot of the deflection angle. This
expression is, of course,  too complicated to use in Eq.
(\ref{Small angles lens equation}). However, we can see that
$\alpha$ diverges when $x_0\rightarrow 3/2$. On the other hand, we
are interested just into small closest approaches, since they
correspond to the high deflection angles producing relativistic
images. If we let $x_0=3/2+\epsilon$, we can search for the
leading order term in the divergence when $\epsilon \rightarrow
0$.

After some non--trivial expansions, we find that the leading order
of the deflection angle is logarithmic in $\epsilon$, that is
\begin{equation}
\alpha \sim -2 \log \frac{\left(2+\sqrt{3} \right) \epsilon}{18
}-\pi .%
\label{Expanded deflection angle}
\end{equation}

\begin{figure}
\resizebox{\hsize}{!}{\includegraphics{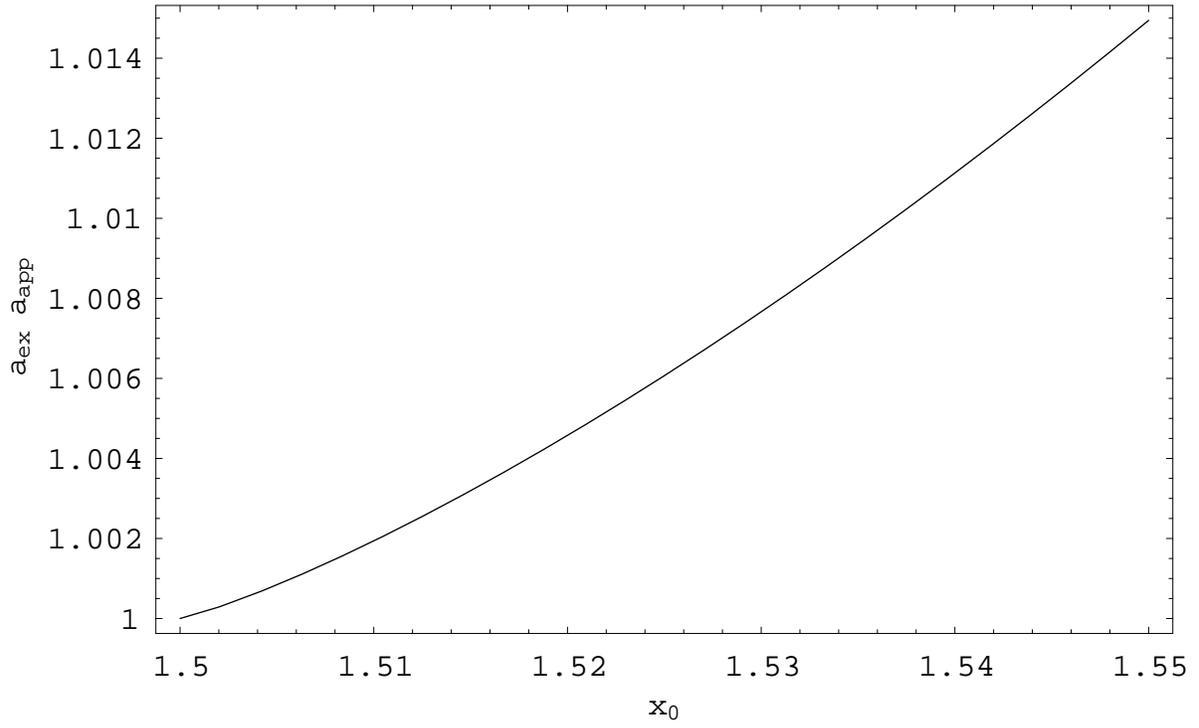}}
 \caption{Ratio of the exact deflection angle %
 (\ref{Deflection angle}) and the approximate one %
 (\ref{Expanded deflection angle}) as functions of the closest approach %
 distance.}
 \label{F3}
\end{figure}

In Fig. \ref{F3}, we plot the ratio of the exact deflection angle
and its leading order expansion. Of course, the two functions
coincide in the limit $x_0 \rightarrow 1.5$. Using our expansion
up to $x_0=1.55$, the error we commit is about $1\%$. For our
purposes, this is largely sufficient as we shall see later.

The next step is to convert the dependence on $x_0$ into a
dependence on the impact parameter $b$ and then on $\theta$. The
study of light ray dynamics in Schwarzschild metric gives the
relation \cite{Wei}
\begin{equation}
x_0^2=\left(1-\frac{1}{x_0} \right) b^2 .
\end{equation}
For values of $x_0$ close to $3/2$, this relation can be solved
perturbatively (or, equivalently, one can expand the exact root of
the third degree polynomial equation in $x_0$ to the first
non-trivial order), giving the relation
\begin{equation}
\epsilon=\sqrt{\frac{b-\frac{3\sqrt{3}}{2}}{\sqrt{3}}} .%
\label{Imppar-mindis}
\end{equation}

Remembering that $b\simeq \theta D_{OL}$, we can substitute this
relation into Eq. (\ref{Expanded deflection angle}) to retrieve
$\alpha$ as a function of $\theta$, that is
\begin{equation}
\alpha \sim -\log\left( \theta D_{OL}-\frac{3\sqrt{3}}{2}
\right)+A ,%
\label{Final deflection angle}
\end{equation}
with
\begin{equation}
A=-\log\frac{\left(5+3\sqrt{3} \right)}{1944}-\pi=2.109 .
\end{equation}

Eq. (\ref{Final deflection angle}) is the highly simplified
expression for the deflection angle we were looking for. It allows
reliable calculations, as we shall see in the next section, where
we insert it in the lens equation.

\section{Images, magnification, critical curves}

According to the considerations done at the end of Sect. 2, what
enters the lens equation is not the full deflection angle but its
deviation from a multiple of $2\pi$.  First of all, we have to
find the values of $\theta$ (denoted by $\theta_n^0$) such that
\begin{equation}
\alpha \left( \theta_n^0 \right) =2n\pi .
\end{equation}
Inverting this equation, with $\alpha$ given by (\ref{Final
deflection angle}), we have
\begin{equation}
\theta_n^0=\frac{3\sqrt{3}+2e^{A-2n\pi}}{2D_{OL}} .%
\label{Theta0}
\end{equation}

As expected, if we let the number of loops around the black hole
tend to infinity, we have ${\displaystyle \theta_\infty^0 =
\frac{3\sqrt{3}}{2D_{OL}}}$.  $\theta_n^0$ are the starting points
for all the successive calculations. The offsets $\Delta
\alpha_n$ can be found by expanding $\alpha$ to the first order
in $\Delta \theta_n=\theta-\theta_n^0$:
\begin{equation}
\Delta \alpha_n=-\frac{D_{OL}}{e^{A-2n\pi}}\Delta \theta_n .%
\label{Delta alpha}
\end{equation}

We have all the ingredients to solve the lens equation (\ref{Small
angles lens equation}), which becomes
\begin{equation}
\beta=\left( \theta_n^0+\Delta \theta_n \right) + \left(
\frac{D_{OL}}{e^{A-2n\pi}}\frac{D_{LS}}{D_{OS}} \right) \Delta
\theta_n .%
\label{DTheta lens equation}
\end{equation}
We can observe that $D_{OL}$, $D_{LS}$, $D_{OS}$ are all much
greater than unity (remember that all distances are measured in
Schwarzschild radii) and $e^{A-2\pi}=3.25\times 10^{-3}$. This
means that the last term in Eq.(\ref{DTheta lens equation})
prevails on the $\Delta \theta_n$ at the second place in the rhs.
Neglecting this term, we finally get the position of the $n^{th}$
image as
\begin{equation}
\theta_n \simeq \theta_n^0+\frac{e^{A-2n\pi}\left(\beta-\theta_n^0
\right) D_{OS}}{D_{LS}D_{OL}} .%
\label{Images}
\end{equation}

We see that, when $\beta$ equals $\theta_n^0$, there is no
correction to the position of the $n^{th}$ image, that remains in
$\theta_n^0$ simply. In this particular case, the image position
coincides with the source position. It is worthwhile to note that
the second term in (\ref{Images}) is much smaller than the first
one. For practical purposes,  $\theta_n^0$ are already a  good
approximation for the position of relativistic images.

The relation (\ref{Delta alpha}) can help us to estimate the error
in the determination of the position of the images. We  do this in
the least favorable case that is for the first image, which is
the most external and thus the farthest from the divergence we
started from. The relative error for $\theta_1^0$ is
\begin{equation}
\frac{\Delta \theta_1}{\theta_1^0}= \left(
\frac{e^{A-2\pi}}{D_{OL}} \frac{\alpha}{\theta_1^0} \right)
\frac{\Delta \alpha}{\alpha}
\end{equation}
For the first image we have $\alpha \simeq 2\pi$. In the previous
section, we have estimated the relative error on $\alpha$ to be
about one percent for $x_0 \simeq 1.55$. By using $b=\theta
D_{OL}$ and Eq. (\ref{Imppar-mindis}), we can find the closest
approach distance for the first image to be $1.545$. In
conclusion, this estimate for $\Delta \alpha/\alpha$ can be
reasonably used.

The error on $\theta_1^0$ is
\begin{equation}
\frac{\Delta \theta_1}{\theta_1^0} \simeq \frac{4\pi
e^{A-2\pi}}{3\sqrt{3}+2e^{A-2\pi}} \frac{\Delta \alpha}{\alpha}
\simeq 8 \times 10^{-5} .
\end{equation}

For the other images, the errors can be found to be orders of
magnitude smaller. This estimate supports our results which thus
prove to be extremely reliable.

The critical curves are defined as the points where the Jacobian
determinant of the lens equation vanishes, that is
\begin{equation}
\frac{\beta}{\theta}\frac{\partial \beta}{\partial \theta}=0 .
\end{equation}
The term ${\displaystyle \frac{\partial \beta}{\partial \theta}}$
is always positive (there are no radial critical curves). Then the
(tangential) critical curves are obtained for $\beta=0$. As we
have already solved the lens equation for each $\beta$, it is
sufficient to put $\beta=0$ in Eq.(\ref{Images}) to get the angles
of these relativistic curves, that is
\begin{equation}
\theta_n^{{cr}} \simeq \theta_n^0\left(1-\frac{e^{A-2n\pi}D_{OS}
}{D_{LS}D_{OL}}\right) .%
\label{Critical curves}
\end{equation}

A  source perfectly aligned  to the black hole  produces an
infinite series of concentric rings with these radii.

The magnification of the images (\ref{Images}) is nothing else but
the inverse of the modulus of the Jacobian determinant already
used for the critical curves. In the region of our interest, i.e.
for small source angles, it is easy to calculate all the needed
quantities from the lens equation. We do this by approximating the
images by the angles $\theta_n^0$:
\begin{equation}
\left. \frac{\partial \beta}{\partial \theta}\right|_{\theta_n^0}
=1 +\frac{D_{OL}}{e^{A-2n\pi}}\frac{D_{LS}}{D_{OS}} .
\end{equation}
The second term is much higher than $1$ for the same reasons
expressed before. Then the magnification of the $n^{th}$ image is
\begin{equation}
\mu_n=\frac{1}{\left| detJ|_{\theta_n^0}\right|}=
\frac{\theta_n^0}{\beta \left. \frac{\partial \beta}{\partial
\theta}\right|_{\theta_n^0}} =e^{A-2n\pi}\frac{\left(
3\sqrt{3}+2e^{A-2n\pi}\right)
D_{OS}}{2\beta D_{OL}^2 D_{LS}} .%
\label{Magnification}
\end{equation}

This expression gives a magnification decreasing very quickly and
then the luminosity of the first image dominates all the others.
The amplification diverges for $\beta \rightarrow 0$, confirming
the fact that the possible detection of relativistic images is
maximal for a perfect alignment of the source with the lens. In
our approximation, the amplification of the images on the opposite
side of the source is just the same of those on the same side.

Another interesting quantity is the total magnification of the
relativistic images. As these would be seldom resolved as single
images, it is likely to see them as one image with a total flux
equal to the sum of the partial contributions coming from each
image. Then, we just have to sum up the series
\begin{equation}
\mu_{{tot}}=2\sum\limits_{n=1}^\infty \mu_n ,
\end{equation}
which is a geometrical one, and then
\begin{equation}
\mu_{{tot}}=\frac{e^A \left(3\sqrt{3} \left(1+e^{2\pi}
\right)+2e^A \right) D_{OS} }{\left( e^{4\pi}-1 \right)\beta
D_{OL}^2D_{LS}} . \label{Total magnification}
\end{equation}

To take into account the finite extension of the source, one must
integrate over its luminosity profile. As a simple example,
integrating the $\frac{1}{\beta}$ dependence in Eqs.
(\ref{Magnification}) and (\ref{Total magnification}) over a
uniform disk with angular radius $\beta_{S}$, we get
\begin{eqnarray}
\frac{1}{\pi \beta_{S}^2}\int\limits_{D(\beta,\beta_{S})}
\frac{1}{\beta'}{d}^2 \beta'&=&   \frac{2 Sign
\left[\beta_{S}-\beta\right]}{\pi \beta_{S}^2} \left[ \left(
\beta_{S}-\beta \right) E \left(\frac{\pi}{2}, -\frac{4 \beta_{S}
\beta}{\left( \beta_{S}-\beta \right)^2} \right)+ \right.\label{Extended integral}\\
&+&\left. \left( \beta_{S}+\beta \right) F \left(\frac{\pi}{2},
-\frac{4 \beta_{S} \beta}{\left( \beta_{S}-\beta \right)^2}
\right)\right] ,\nonumber
\end{eqnarray}
where we have indicated the disk with radius $\beta_{S}$ centered
on $\beta$ by $D(\beta,\beta_{S})$ and
\begin{equation}
E \left( \phi_0,\lambda
\right)=\int\limits_0^{\phi_0}\left(1-\lambda \sin^2 \phi
\right)^{1/2} {d} \phi ,%
 \label{Elliptic function II}
\end{equation}
is the elliptic integral of the second kind.

Then, the magnification of an extended uniform source of angular
radius $\beta_{S}$ can be obtained by substituting the $1/\beta$
dependence in Eq. (\ref{Magnification}) and Eq. (\ref{Total
magnification}) with the right member of Eq. (\ref{Extended
integral}).

\section{Discussion}

The formulae just derived provide a complete characterization of
the two infinite sets of relativistic images surrounding a black
hole or, in general,   any compact object acting as a lens whose
size is comparable with its Schwarzschild radius. They can be
employed in all kinds of phenomenological calculations to test
their detectability.

Some general considerations can be done on the features expected
by a candidate lensing system. First, we need a very compact
massive object, possibly a black hole, in order to gain access to
the region of strong field immediately outside of the event
horizon. Moreover, the matter surrounding this compact object
should be transparent to the wavelength of the radiation emitted
by the lensed source, otherwise the photons would be  absorbed
before the light rays complete their loops and would not reach
the observer.

We recall that the amplification of each of the {\it weak field
images} can be expressed (with distances in Schwarzschild units)
as:
\begin{equation}
\mu_{{wfi}}=\frac{1}{\beta}\sqrt{\frac{2D_{LS}}{D_{OL}D_{OS}}}
\end{equation}
for the small $\beta_s$ we are interested in. We can observe that
the dependence on $\beta$ is the same of the weak field images.
Then the relative importance of the relativistic images to the
weak field ones is constant for high alignments of the source to
the lens.

The ratio of this quantity and the total magnification of the
relativistic images (\ref{Total magnification}) is
\begin{equation}
\frac{\mu_{{wfi}}}{\mu_{{tot}}}=\frac{\sqrt{2}}{B} \left(
\frac{D_{OL}D_{LS}}{D_{OS}}\right)^{3/2}%
\label{Magnification ratio}
\end{equation}
where $B$ is a numerical coefficient:
\begin{equation}
B=\frac{e^A \left(3\sqrt{3} \left(1+e^{2\pi} \right)+2e^A \right)
}{\left( e^{4\pi}-1 \right)}=0.017
\end{equation}

Then we must expect relativistic images to be always very faint
with respect to the weak field images, since this ratio goes as an
astronomical distance elevated to $3/2$. This fact rules out
microlensing as a method for detection of relativistic images.

The separation between the two sets of relativistic images is of
the order of the Schwarzschild diameter of the compact object.
Depending on the specific situation, the angle corresponding to
this length is generally very small. A very massive black hole
would surely help to separate these images each other. The mass
of the black hole at the center of our Galaxy \cite{Ric} is
believed to be about $2.8 \times 10^{6}M_\odot$. Its
Schwarzschild angle would be about $15$ microsecs, which could
become accessible by VLBI experiments \cite{Hir,Ulv}. In this
case, at least the two sets of relativistic images of some rear
source would be distinguishable. With their very low
magnification, depending on the alignment degree, the
relativistic images of any rear source would cast a very hard but
not impossible challenge for observational astronomy.

Possible better candidates could be black holes at the centers of
other galaxies lensing some compact source (e.g. Quasars) on the
background, or black holes at the centers of globular clusters.

\section{Conclusions}

The expansion we have presented in this paper can be considered as
the strong field gravitational lensing limit, in opposition to the
usual weak field limit. Its starting point is the critical null
geodesic followed by a massless particle captured by the black
hole. We have expanded the deflection angle in this limit and used
its leading order in the full lens equation. We have solved this
equation and found two infinite sets of relativistic images. We
have given a complete analytical description of these images with
simple formulae for their position, magnification and for the
critical curves.

The striking importance of these images lies on the fact that
they could provide a profound test of general relativity in its
full regime. In this situation, it would be surely possible to
distinguish among relativistic theories of gravitation (e.g.
Brans-Dicke, induced gravity, etc). For this reason, it is
imperative to perform similar expansions within alternative
pictures and evaluate the differences with the standard results
we have found.

The scheme we have used for this calculation encourages its
application to more complicated problems of general relativity
such as charged rotating black holes or other relevant
astrophysical objects where strong field general relativity is
involved.

\begin{centerline}
{\bf Acknowledgments}
\end{centerline}

Work supported by fund 60\% D.P.R. 382/80.

\newpage

\newpage

\newpage

\end{document}